\documentclass[12pt]{amsart}

\newtheorem%
{thm}{Theorem}[section]
\newtheorem%
{proposition}[thm]{Proposition}
\newtheorem%
{lemma}[thm]{Lemma}
\newtheorem%
{lemmadef}[thm]{Lemma-Definition}
\newtheorem%
{corollary}[thm]{Corollary}
\newtheorem%
{conjecture}[thm]{Conjecture}

\newcommand{\dontprint}[1]{\relax}

\def\gh{\mathrm{gh}}
\title
[Schwinger-Dyson equation for non-Lagrangian field theory]
{Schwinger-Dyson equation \\[3mm] for non-Lagrangian field theory}
\author{ S.L. Lyakhovich and  A.A. Sharapov}
\address{Department of Quantum Field Theory, Tomsk State University, Tomsk 634050, Russia}
\email{sll@phys.tsu.ru, sharapov@phys.tsu.ru}

\thanks{We are thankful to Jim Stasheff for his corrections to and comments on the first version of the manuscript. This work was
partially supported by the RFBR grant 06-02-17352, Russian
Ministry of Education grant 130337, and the grant for Support of
Russian Scientific Schools 1743.2003.2. AAS appreciates financial
support from the Dynasty Foundation and International Center for
Fundamental Physics in Moscow.}

\begin{document}

\begin{abstract}

A method is proposed of constructing quantum correlators for a
general gauge system whose classical equations of motion do not
necessarily follow from the least action principle. The idea of
the method is in assigning a certain BRST operator $\hat\Omega$ to
any classical equations of motion, Lagrangian or not. The
generating functional of Green's functions is defined by the
equation $\hat\Omega Z (J) = 0$ that is reduced to the standard
Schwinger-Dyson equation whenever the classical field equations
are Lagrangian. The corresponding probability amplitude $\Psi$ of a
field $\varphi$ is defined by the same equation $\hat\Omega \Psi
(\varphi) = 0$ although in another representation. When the
classical dynamics are Lagrangian, the solution for $\Psi
(\varphi)$ is reduced to the Feynman amplitude
$e^{\frac{i}{\hbar}S}$, while in the non-Lagrangian case this
amplitude can be a more general distribution.
\end{abstract}

\maketitle


\section{Introduction}
In this paper we introduce a generalization of the Schwinger-Dyson
equation for dynamical systems whose classical equations of motion
are not required to follow from the least action principle. The
generating functional of Green's functions, being a solution of
the generalized Schwinger-Dyson equation, defines quantum
correlators in non-Lagrangian field theory.

It has been recently shown \cite{KLS} that any classical field
theory in $d$ dimensions, be it Lagrangian or not, can be
converted into an equivalent Lagrangian topological field theory
(TFT) in $d+1$ dimension. Path integral quantizing this effective
Lagrangian TFT, one gets correlators for the original
non-Lagrangian theory in $d$ dimensions. If the original field
theory admits an action principle, the path integral for the
enveloping $(d+1)$-dimensional TFT can be explicitly integrated
out in the bulk with the result that reproduces the standard
Batalin-Vilkovisky quantization receipt \cite{BV}, \cite{HT} for
the original Lagrangian gauge theory. Thus, at least one
systematic method is known of constructing quantum correlators for
general non-Lagrangian systems. In the present paper, we formulate
a generalization of the Schwinger-Dyson equation for
non-Lagrangian theories that defines Green's functions in terms of
the original space, i.e., without recourse to embedding into a
Lagrangian TFT in $d+1$ dimensions. Below in the introduction, we
give some elementary explanations of the main idea behind our
construction. More rigorous and detailed exposition is given in
the subsequent sections.

Consider the dynamics of fields $\phi=\{\phi^i\}$ governed by the
equations of motion\footnote{Hereinafter we use De Witt's
condensed notation \cite{DW}, whereby the superindex  ``$i$''
comprises both the local coordinates on a $d$-dimensional
space-time manifold $M$ and possible discrete indices labelling
different components of $\phi$. As usual, the repeated
superindices, e.g. $\phi^i\psi_i$, imply summation over the
discrete indices and integration over the space-time coordinates
w.r.t. an appropriate measure on $M$. The partial derivatives
$\partial_i=\partial/\partial \phi^i$ are understood as
variational ones.}
\begin{equation}\label{T1}
    T_a(\phi)=0\,.
\end{equation}
These equations are not assumed to come from the least action
principle, so the indices ``$i$'' and ``$a$'', labelling the
fields and the equations, may belong to different sets completely
unrelated to each other. For a Lagrangian theory, however, these
indices must coincide because
\begin{equation}
 T_i(\phi)=\frac{\partial S(\phi)}{\partial\phi^i}\,,
 \label{TL}
 \end{equation}
where $S$ is an action functional. The standard Schwinger-Dyson
equation \cite{Schwinger} for the generating functional  of
Green's functions $Z(J)$ reads
\begin{equation}
\hat{{\mathbb{T}}}_i Z(J) =0, \qquad \hat{{\mathbb{T}}}_i = - J_i
+ T_i (\phi)|_{\phi \mapsto i\hbar \frac{\partial}{\partial J}}\,.
\label{SDE}
\end{equation}
To avoid further restrictions on $Z(J)$, the operators
$\hat{{\mathbb{T}}}_i$ must commute with each other, which
requires the equations of motion to be Lagrangian:
\begin{equation}
[\hat{{\mathbb{T}}}_i , \hat{{\mathbb{T}}}_j] =  i\hbar (
\partial_i {T}_j -
\partial_j T_i)|_{\phi \mapsto i\hbar
\frac{\partial}{\partial J}}= 0 \quad \Leftrightarrow \quad
T_i=\partial_i S\,,\label{A}
\end{equation}
for some action $S$.

One can regard the Schwinger-Dyson operators
$\hat{{\mathbb{T}}}_i$ as those resulting from the canonical
quantization of the abelian first-class  constraints
\begin{equation}
{{\mathbb{T}}}_i (\phi, J) =  T_i(\phi)-J_i\, , \qquad \{
{{\mathbb{T}}}_i , {{\mathbb{T}}}_j \} = 0\,.
 \label{Aconstr}
 \end{equation}
Here we consider the sources $J_i$ as the momenta canonically
conjugate to the fields $\phi^i$,
\begin{equation}\label{PBJ}
\{ \phi^i ,  \phi^j \}=0\,,\qquad\{ \phi^i , J_j \} = \delta^i_j
\, , \qquad \{ J_i , J_j \}=0 \, ,
\end{equation}
and use the momentum representation to pass from functions to
operators:
\begin{equation}
\hat{J}_i=J_i\cdot\,,\qquad \hat{\phi}{}^i= i\hbar
\frac{\partial}{\partial J^i}\,.
\end{equation}

Since the constraints (\ref{Aconstr}) are explicitly solved w.r.t.
the momenta $J$, they must be  abelian  whenever they are first
class. So, the property of the equations of motion to be
Lagrangian is equivalent to the property of the constraints
(\ref{Aconstr}) to be  first class. One can also quantize the
constraints (\ref{Aconstr}) in the coordinate representation
related to the momentum one by the Fourier transform:
\begin{equation}\label{Coord}
\hat{{\mathbb{T}}}_i =  T_i (\phi) +
i\hbar\frac{\partial}{\partial \phi^i} \, .
\end{equation}
Imposing this constraint operator on a ``state" $\Psi(\phi)$ which
is the Fourier transform of the generating functional $Z(J)$,
yields
\begin{equation}\label{FA}
\left[ T_i (\phi) \, + \, i\hbar\frac{\partial}{\partial \phi^i}
\right] \Psi(\phi) = 0\,.
\end{equation}
For Lagrangian equations of motion (\ref{TL}), the solution is
given by the Feynman probability amplitude
\begin{equation}\label{PA}
\Psi(\phi) =(\mathrm{const}) e^{\frac{i}{\hbar}S(\phi)}\,.
\end{equation}
Thus, from the viewpoint of the phase space (\ref{PBJ}) of fields
and sources, the Schwinger-Dyson equation is nothing but imposing
the quantum first-class constraints (\ref{Aconstr}) on the
probability amplitude. Given the generating functional or
corresponding probability amplitude, the quantum average of an
observable $\mathcal{O}(\phi)$ is given by the path integral
\begin{equation}\label{Qaver}
\langle \mathcal{O}\rangle = \int D\phi\, \mathcal{O}(\phi)
\Psi(\phi) = \mathcal{O}\left(i \hbar\frac{\partial}{\partial
J}\right) Z(J)|_{J=0}\, ,
\end{equation}
where $D\phi$ is an integration measure on the configuration space
of fields.

Notice that the constraints (\ref{Aconstr}) start with the
classical equations of motion $T(\phi)$ which are always Poisson
commuting w.r.t. (\ref{PBJ}), be the equations Lagrangian or not.
It is a common practice to regard the equations of motion as
constraints in the space of fields, see e.g. \cite{HT}, so it
might be interesting to impose the classical equations themselves
on the probability amplitude without adding the momentum term $J$
to $T$:
\begin{equation}\label{ClSDE}
    T(\phi) \Psi (\phi) = 0\,.
\end{equation}
The solution is obvious,
\begin{equation}\label{ClA}
\Psi (\phi) \sim \delta(T(\phi))\,.
\end{equation}
The function $\Psi (\phi)$  is the classical probability amplitude
in the sense that the quantum average of any observable
$\mathcal{O}(\phi)$ would be proportional to its classical
on-shell value when calculated with this amplitude:
\begin{equation}\label{Claver}
\langle \mathcal{O}\rangle = (\mathrm{const})\int \mathcal{D}\phi
\,\delta(T(\phi)) \mathcal{O}(\phi) \sim \mathcal{O}(\phi_0)\,,
\end{equation}
$\phi_0$ being a solution to the classical field equations
(\ref{T1}). The classical probability amplitude (\ref{ClA}) was
introduced and studied by Gozzi et al \cite{Gozzietall} for
Hamiltonian equations of motion. It was also shown in
\cite{Gozzietall} that (\ref{ClA}) is a classical limit of
Feynman's amplitude (\ref{PA}).

A twofold general conclusion can be derived from these simple
observations. First, taking the classical limit for the
Schwinger-Dyson equation means the omission of the momentum term
in the first class constraints (\ref{Aconstr}). To put this
another way, quantizing means extending the classical equations of
motion, viewed as constraint operators imposed on the probability
amplitude, by appropriate momentum terms so that the constraints
remain first class. The second simple lesson is that the classical
equations of motion, being viewed as quantum constraints, are
first class, be the original equations Lagrangian or not.
Regarding these observations, the way seems straightforward of
constructing the generalization of the Schwinger-Dyson equation
for non-Lagrangian classical systems: The Schwinger-Dyson
operators are to be constructed by extending the general equations
(\ref{T1}) with momentum terms. These momentum terms are to be
sought for from the condition that the momentum-extended
constraints must be  first class.

Proceeding from the heuristic arguments above, we can take the
following ansatz for the $\phi J$-symbols of the Schwinger-Dyson
operators:
\begin{equation}\label{TT}
{{\mathbb{T}}}_a(\phi,J)=T_a(\phi)+V_a^i(\phi)J_i+ O(J^2)\,.
\end{equation}
These operators are defined as formal power series in momenta
(sources) $J$ with leading terms being the classical equations of
motion. Requiring the Hamiltonian constraints $\textsc{T}_a=0$ to
be  first class, i.e.,
\begin{equation}\label{inv}
    \{{\mathbb{T}}_a, {\mathbb{T}}_b\}=U_{ab}^c {\mathbb{T}}_c \,,\qquad
    U_{ab}^c(\phi,J)=C^c_{ab}(\phi)+O(J)\,,
\end{equation}
one gets an infinite set of relations on the structure
coefficients in the expansion of ${{\mathbb{T}}}_a(\phi,J)$ in
$J$. In particular, examining the involution relations (\ref{inv})
to zero order in $J$ we get
\begin{equation}\label{anchor}
    V_a^i\partial_iT_b-V_b^i\partial_i T_a=C_{ab}^cT_c\,.
\end{equation}
The value $V_a^i(\phi)$ defined by this relation is called the
\textit{{Lagrange anchor}}. It has been first introduced in the
work \cite{KLS} with a quite different motivation. The Lagrange
anchor is a key geometric ingredient for converting a
non-Lagrangian field theory in $d$ dimensions into an equivalent
Lagrangian TFT in $d+1$ dimensions  \cite{KLS}.

If the field equations (\ref{T1}) are Lagrangian, one can choose
the Lagrange anchor to be  $-\delta^i_j$.  This choice results in
the standard Schwinger-Dyson operators (\ref{SDE},\ref{Aconstr})
having abelian involution (\ref{A}). For general equations of
motion (\ref{T1}), the Lagrange anchor has to be field-dependent
(that implies non-abelian involution (\ref{inv}), in general)  and
is not necessarily invertible. Existence of the invertible
Lagrange anchor is equivalent to existence of a Lagrangian for the
equations (\ref{T1}). Zero anchor ($V=0$) is admissible for any
equations of motion, Lagrangian or non-Lagrangian, as it obviously
satisfies (\ref{anchor}). The zero anchor, however, leads to a
pure classical equation (\ref{ClSDE}) for the probability
amplitude, and no quantum fluctuations would arise in such theory.
Any nonzero Lagrange anchor, invertible or not, allows one to
construct the generating functional of Green functions describing
nontrivial quantum fluctuations.

Given the Lagrange anchor,  the requirement for the symbols of the
Schwinger-Dyson operators (\ref{TT}) to be the first class
constraints (\ref{inv}) will recursively  define all the higher
terms in (\ref{TT}) as we explain in the next section. Since the
general Lagrange anchor results in the non-abelian algebra
(\ref{inv}) of the Schwinger-Dyson operators, the naive imposing
of these operators on the generating functional $Z(J)$ or
probability amplitude $\Psi (\phi)$
\begin{equation}\label{GSDE}
\hat{{\mathbb{T}}}_a \Psi(\phi)=0
\end{equation}
could be not a self-consistent procedure. As always with
non-abelian constraints, the BFV-BRST formalism \cite{BFV},
\cite{HT} gives the  most systematic method to handle all the
consistency conditions of the constraint algebra. Instead of
naively imposing constraint operators, the BFV-BRST method implies
seeking for the probability amplitudes defined as cohomology
classes of the nilpotent operator associated to the first class
constraints. So, constructing the BRST operator for the
Schwinger-Dyson constraints (\ref{TT}), (\ref{inv}), we will get
the equation defining quantum correlators for the non-Lagrangian
theory. Notice that for the Lagrangian field theory, it was  known
long ago \cite{AD}, \cite{HT} that the Schwinger-Dyson equation
can be reinterpreted as the Ward identity for an appropriately
introduced BRST symmetry. This idea, as we will see, still allows
the derivation of the Schwinger-Dyson equation for non-Lagrangian
field theory by making use of the BRST symmetry related with the
Langange anchor.

The paper is organized as follows. In Sect. 2, we recall some
results from Ref. \cite{KLS}. In particular, we describe the
classical BRST complex one can associate with any (non)-Lagrangian
gauge system. We also explain how the classical observables are
described as BRST cohomology of this complex. In Sect. 3, this
BRST complex is used to construct the probability amplitude on the
configuration space of fields. We also show that the equation
defining this amplitude is reduced to the standard Schwinger-Dyson
equation whenever the theory is Lagrangian. Sect. 4 gives the path
integral representation for the probability amplitude and the
quantum averages of physical observables. In Sect. 5  we compare
the proposed quantization scheme for (non)-Lagrangian gauge
theories with the standard BV-quantization based on Lagrangian
formalism. In the Lagrangian case, our method reproduces the BV
scheme although some basic properties of quantum theory can have a
more general form whenever no action principle is admissible for
the classical dynamics. In particular, in the general
non-Lagrangian case, the classical BRST differential can be a
non-inner derivation of the antibracket and the probability
amplitude is not necessarily an exponential of the master action.

\section{Equations of motion as phase-space constraints }

Consider a collection of fields $\phi=\{\phi^i\}$ subject to
(differential) equations of motion
\begin{equation}\label{t}
    T_a(\phi)=0
\end{equation}
and an admissible set of boundary conditions.  For the sake of
simplicity, assume that the fields $\phi$ are bosons; fermion
fields and equations of motion can be easy incorporated into the
formalism by inserting appropriate sign factors in the subsequent
formulas.

It is convenient to think of $\phi$ as a coordinate system  on an
(infinite-dimensional) manifold $\mathcal{M}$ of all field
configurations with prescribed boundary conditions; upon this
interpretation one can regard $T=\{T_a(\phi)\}$ as a section of
some vector bundle $\mathcal{E}\rightarrow \mathcal{M}$ over the
base $\mathcal{M}$. The set $\phi_0=\{\phi_0^i\}\subset
\mathcal{M}$ of all solutions to Eqs.(\ref{t}) coincides then with
the zero locus of the section $T\in \Gamma(\mathcal{E})$. Under
the standard regularity conditions \cite{HT},  $\{\phi_0\}\subset
\mathcal{M}$ is a smooth submanifold associated with an orbit of
gauge symmetry transformations (see Eq.(\ref{RT}) below); for
non-gauge invariant systems this orbit consists of one point.

Now let $T^\ast \mathcal{M}$ be a cotangent bundle of the space
$\mathcal{M}$, with $\bar\phi=\{\bar\phi_i\}$ being the fiber
coordinates. Using the physical terminology, we will refer to
$\bar\phi$ as the momenta conjugate to the ``position
coordinates'' $\phi$. The canonical Poisson brackets on $T^\ast
\mathcal{M}$ read
\begin{equation}\label{pb}
   \{\phi^i,\phi^j\}=0\,,\quad \{\phi^i,\bar\phi_j\}=\delta_j^i \,,\quad
    \{\bar\phi_i,\bar\phi_j\}=0\,.
\end{equation}
We can regard the original equations of motion (\ref{t}) as
holonomic constraints on the phase space $T^\ast \mathcal{M}$. \
\/From this viewpoint, the linearly dependent equations of motion
correspond to reducible Hamiltonian constraints with Noether's
identity generators $Z=\{Z_A^a(\phi)\}$ playing the role of
null-vectors for the constraints,
\begin{equation}\label{ZT}
    Z_A^aT_a=0\,.
\end{equation}
It may also happen that Eqs. (\ref{t}) are gauge invariant (i.e.,
they do not specify a unique solution), in which case there exists
a set of nontrivial gauge symmetry generators $R=\{R_\alpha^i\}$
such that
\begin{equation}\label{RT}
    R_\alpha^i\partial_iT_a=U_{\alpha a}^bT_b\,,
\end{equation}
for some structure functions $U^b_{\alpha a}(\phi)$.  Then we can
enlarge the set of holonomic constraints $T_a=0$ by the
constraints
\begin{equation}\label{Rconstr}
R_\alpha =R_\alpha^i(\phi)\bar\phi_i
\end{equation}
that are linear in momenta.

Assume that the Noether identity generators and the gauge symmetry
generators are complete and irreducible in the usual sense
\cite{HT}. Then, in consequence  of (\ref{RT}) and completeness of
the generators $R$, the combined set of constraints
$\Theta_I=(T_a, R_\alpha)$ is first class:
\begin{equation}\label{ThetaInv}
\{\Theta_I,\Theta_J\}=U_{IJ}^K\Theta_K\quad\Leftrightarrow\quad
\begin{array}{l}
\{T_a,T_b\}=0\,,\\[3mm]
\{R_\alpha, T_a\}=U_{\alpha a}^bT_b\,,\\[3mm]
\{R_\alpha,R_\beta\}=W_{\alpha\beta}^\gamma R_\gamma +T_a
E^{ai}_{\alpha\beta}\bar\phi_i\,,
\end{array}
\end{equation}
$W_{\alpha\beta}^\gamma(\phi)$, $E^{ai}_{\alpha\beta}(\phi)$ being
some structure functions. In view of (\ref{ZT}) the constraints
$\Theta$  are reducible
\begin{equation}\label{Xi}
\Xi_A^I\Theta_I = 0 \, , \qquad  \Xi_A^I = (Z_A^a,0)\,.
\end{equation}
According to the terminology of \cite{KLS}, Rels. (\ref{ZT}),
(\ref{RT}) define a gauge theory of type (1,1). Notice that the
generators of gauge symmetry $R$ and generators of Noether's
identities $Z$ can be completely independent from each other in
the case of non-Lagrangian theory \cite{KLS}. In particular, it is
possible to have  gauge invariant but linearly independent
equations of motion, and conversely, a theory may have linearly
dependent equations without gauge invariance. In these cases we
speak about gauge theories of type (1,0) and (0,1), respectively.
The theories of type (0,0) are those for which  Eqs. (\ref{t}) are
independent and have a unique solution. The general ($n,m$)-type
gauge theory with $n>1$ and/or $m>1$ corresponds to the case of
$(n-1)$-times reducible generators of gauge transformations and
$(m-1)$-times reducible generators of the Noether identities (see
\cite{KLS} for the precise definition).

It is easy to see that the ``number" of independent first class
constraints among the $\Theta$'s coincides with the ``number" of fields
$\phi$. The same can be said in a more formal way: the coisotropic
surface $\mathcal{L}\subset T^\ast \mathcal{M}$ defined by the
first-class constraints (\ref{ThetaInv}) is a Lagrangian
submanifold. Consider the action of the Hamiltonian system with
constraints $\Theta$:
\begin{equation}\label{top}
    S[\lambda,\phi,\bar\phi]=\int_{t_1}^{t_2}dt(\bar\phi_i\dot{\phi}^i
    -\lambda^I\Theta_I(\phi,\bar\phi))\,.
\end{equation}
This action corresponds to a purely topological field theory
 having no physical evolution w.r.t. to the time \footnote{Notice, that the ``time" $t$ in (\ref{top})
is an auxiliary $(d+1)$-st dimension, which is not related anyhow
with the evolution parameter in the (differential) equations of
motion (\ref{t}) . The true physical time is among the $d$ original
dimensions.} $t$. The model is invariant under the standard gauge
transformations generated by the first class constraints
(\ref{ThetaInv}) and their null-vectors (\ref{Xi}):
\begin{equation}\label{g-tran}
\begin{array}{c}
    \delta_{\varepsilon} \phi^i=\{\phi^i,\Theta_I\}\varepsilon^I\,,\qquad \delta_{\varepsilon} \bar
    \phi_i = \{\bar \phi_i, \Theta_I\}\varepsilon^I\,,\\[5mm]
    \delta_{\varepsilon} \lambda^I=\dot{\varepsilon}{}^I -\lambda^K
    {U}_{KJ}^I\varepsilon^J +\Xi^I_A\varepsilon^A\,.
\end{array}
\end{equation}
Here $\varepsilon^I=(\varepsilon^a,\varepsilon^\alpha)$  and
$\varepsilon^A$ are infinitesimal gauge parameters, and the
structure functions ${U}_{KJ}^I(\phi)$ are defined by
(\ref{ThetaInv}).

Imposing the boundary conditions on the momenta
\begin{equation}\label{bcon}
    \bar\phi_i(t_1)=\bar\phi_i(t_2)=0\,,
\end{equation}
one can see that dynamics of the model (\ref{top}) is equivalent
to that described by the original equations (\ref{t}). Indeed, let
$\gamma(t)=(\phi(t), \bar\phi(t), \lambda(t))$ be a trajectory
minimizing the action (\ref{top}). Due to the equations of motion
\begin{equation}\label{constr}
    \frac{\partial S}{\partial \lambda^I}=0\,,
\end{equation}
the phase-space projection $(\phi(t),\bar\phi(t))$ of $\gamma(t)$
lies on the constraint surface $\mathcal{L}$. Given a time moment
$t_0\in (t_1,t_2)$, one can always find an appropriate gauge
transformation (\ref{g-tran}) moving the point
$(\phi(t_0),\bar\phi(t_0))\in \mathcal{L}$ to any other point of
the constraint surface and simultaneously assigning any given
value to $\lambda(t_0)$, no matter how the boundary points
$\gamma(t_1), \gamma(t_2)$ were fixed. So, there are no true
physical dynamics in the bulk $(t_1,t_2)$ and the only nontrivial
equations to solve are the constraints (\ref{constr}) on the
boundary values of fields $(\phi(t),\bar\phi(t))$ at $t_1$ and
$t_2$.  But the boundary condition (\ref{bcon}) reduces Eqs.
(\ref{constr}) to the original equations of motion
$T_a(\phi(t_1))=T_a(\phi(t_2))=0$. Thus, we have two copies of the
original dynamics corresponding to the end points of the ``time"
interval $[t_1,t_2]$.

It is easy to see that the quantization of the topological model
(\ref{top}) induces a trivial quantization of the original theory
\cite{KLS}: the quantum averages of physical observables coincide
exactly with their classical values, i.e., no quantum corrections
arise. In a theory without gauge symmetries and identities, this
is clearly seen from Rels. (\ref{ClSDE}-\ref{Claver}). To get a
nontrivial quantization, one has to modify the model (\ref{top})
keeping intact its classical dynamics. For this end we consider a
formal deformation of the constraints (\ref{ThetaInv}) and their
null-vectors (\ref{Xi}) by higher powers of momenta,
\begin{equation}\label{deform}
\hspace{-2cm}\tilde{\Theta}_I=(\tilde{T}_a,\tilde{R}_\alpha)\,,\qquad\quad
\begin{array}{l}
    \tilde{T}_a=T_a(\phi)+V_a^i(\phi)\bar\phi_i+O(\bar\phi^2)\,,\\[3mm]
    \tilde{R}_\alpha=R_\alpha^i(\phi)\bar\phi_i+O(\bar\phi^2)\,,\\[3mm]
    \tilde{Z}_A^a=Z_A^a(\phi)+O(\bar\phi)\,.
    \end{array}
\end{equation}
In order for the new constraints $\tilde{\Theta}_I$ to define a
topological model - a theory without physical degrees of freedom
in the bulk of the time interval  - they must be first class and
have $\tilde{\Xi}^I_A=(\tilde{Z}_A^a,0)$ as null-vectors. These
requirements result in an infinite set of relations on the
expansion coefficients in (\ref{deform}). In particular, at the
zeroth order in $\bar\phi$'s we get
\begin{equation}\label{VT}
    V_a^i\partial_iT_b-V_b^i\partial_iT_a=C_{ab}^cT_c\,,
\end{equation}
cf. (\ref{TT}-\ref{anchor}). The condition (\ref{VT}) is necessary
and sufficient for the existence of higher order deformations
respecting the algebraic structure of constraints \cite{KLS}. The
functions $(V_a^i,C_{ab}^c)$ are said to define a \textit{Lagrange
structure} compatible with the equations of motion (\ref{T1}); the
set $V=\{V_a^i\}$, determining the first order deformation, is
called the Lagrange anchor. Clearly, the condition (\ref{VT}) does
not specify a unique Lagrange structure. As was shown in
\cite{KLS} any two Lagrange structure are related to each other by
the following transformation:
\begin{equation}\label{VG}
    V_a^i\,\rightarrow\, V_a^i+T_bG^{bi}_a+G_a^\alpha R_\alpha ^i
    +\partial_jT_aG^{ij}\,,\qquad C_{ab}^c\,\rightarrow\,
    C_{ab}^c - G_{[a}^{ci}\partial_iT_{b]}-G^\alpha_{[a}U^c_{\alpha b]}+ G_{ab}^AZ_A^c\, ,
\end{equation}
where $G_{ab}^A$, $ G^{bi}_a$, $G_a^\alpha$, $G^{ij}=G^{ji}$ are
arbitrary functions and the square brackets mean
antisymmetrization in indices $a$, $b$. In particular, one can use
these transformations to generate a nontrivial Lagrange structure
from the trivial one $V=C=0$.

Now consider the Hamiltonian action (\ref{top}) with the
constraints $\tilde{\Theta}$ in place of $\Theta$. We claim that
this replacement does not affect the classical dynamics. Indeed,
repeating the arguments above, we conclude that the classical
dynamics are still localized at the boundary; but condition
(\ref{bcon}) reduces (\ref{constr}) to the original equations of
motion (\ref{t}) for any formal deformation (\ref{deform}). The
classical equivalence, however, is not followed by the quantum
one, and as we will see in the next section, it is the nontrivial
Lagrange anchor $V$ that defines possible ``directions'' of
nontrivial quantum fluctuations near the classical solution.

The BRST-BFV quantization of the Hamiltonian constrained system
implies the extension of the original phase space $T^\ast
\mathcal{M}$ by ghost variables \cite{BFV}, \cite{HT}. To each
first class constraint
$\tilde{\Theta}_I=(\tilde{T}_a,\tilde{R}_\alpha)$ we assign the
ghost field $\mathcal{C}^I=(\bar\eta^a,c^\alpha)$ and the
conjugate ghost momentum
$\bar{\mathcal{P}}_I=(\eta_a,\bar{c}_\alpha)$. Since the
constraints $\tilde{\Theta}$ are supposed to be reducible, we also
introduce the canonically conjugate pairs of ghosts-of-ghosts
$(\bar\xi^A,\xi_A )$. The canonical Poisson structure on the ghost
extended phase-space is defined as follows:
\begin{equation}\label{PB}
    \{\mathcal{C}^I,\bar{\mathcal{P}}_J\}=-\delta_J^I\,,\quad\{\xi_B,\bar\xi^A\}=-\delta_B^A\,.
\end{equation}
The Poisson brackets vanish among other variables with the
exception of ones in the original phase space $T^\ast\mathcal{M}$,
which are left unchanged\footnote{Geometrically, one can regard
the ghost variables as linear coordinates in the fibers of a
$\mathbb{Z}\times \mathbb{Z}_2$-graded vector bundle $E$ over the
configuration space $\mathcal{M}$. Upon this interpretation the
canonical Poisson brackets  on the extended phase space - the
total space of $E$ - are modified by terms involving a linear
connection on $E$ (see \cite{HC}, \cite{LS1}, \cite{KLS}, for more
detailed discussion). Here, however, we consider only the simplest
case where $E\rightarrow \mathcal{M}$ is a trivial vector bundle
with flat connection.}. The Grassman parity and the ghost number
assignments of the new fields are given by
\begin{equation}\label{}
\begin{array}{ll}
\epsilon(\mathcal{C}^I)=\epsilon(\bar{\mathcal{P}}_I)=1\,,&\quad\epsilon
({\bar\xi}^A)=\epsilon(\xi_A)=0\,,\\[3mm]
    \gh (\mathcal{C}^I)=-\gh(\bar{\mathcal{P}}_I)=1\,,&\quad
    \gh(\bar\xi^A)=-\gh(\xi_A)=2\,.
    \end{array}
\end{equation}
Upon the complex conjugation the phase-space variables behave  as
\begin{equation}\label{hermprop}
\phi^\ast=\phi\,,\qquad \bar\phi^\ast=\bar\phi\,,\qquad
\mathcal{C}^\ast=\mathcal{C}\,, \qquad \bar{\mathcal{
    P}}^\ast=-\bar{\mathcal{P}}\,,\qquad \xi^\ast=-\xi\,,\qquad
    \bar\xi^\ast=-\bar\xi\,.
\end{equation}

The  gauge structure of the topological model is completely
encoded by the BRST charge
\begin{equation}\label{om}
    \Omega=\mathcal{C}^I\tilde{\Theta}_I+\bar\xi^A\tilde{\Xi}_A^I\bar{\mathcal{P}}_I+\cdots\,.
\end{equation}
 By definition,
 \begin{equation}\label{star}
    \Omega^\ast=\Omega\,,\qquad \gh(\Omega)=1\,,\qquad \epsilon(\Omega)=1\,,
\end{equation}
and the higher orders of ghost variables in (\ref{om}) are
determined from the master equation
\begin{equation}\label{me}
    \{\Omega,\Omega\}=0\,.
\end{equation}

Denote by $\mathcal{F}$ the Poisson algebra of functions on the
ghost-extended phase space. The generic element of $\mathcal{F}$
is given by formal power series in $\mathcal{C},
\bar{\mathcal{P}}, \xi, \bar\xi$ and $\bar\phi$ with coefficients
being smooth (in any suitable sense) functions of $\phi$. The
adjoint action of $\Omega$ makes $\mathcal{F}$ a cochain complex:
A function $F$ is said to be BRST-closed (or BRST-invariant) if
$\{\Omega, F\}=0$, and a function $B$ is said to be  BRST-exact
(or trivial) if $B=\{\Omega, C\}$, for some $C$. The corresponding
cohomology group $\mathcal{H}=\bigoplus_{k\in
\mathbb{Z}}\mathcal{H}_k$ is naturally graded by the ghost number.

As usual, the physical observables are nontrivial BRST invariants
with ghost number zero. It can be shown \cite{KLS} that the
cohomology class of any BRST-invariant function
$F=F(\phi,\bar\phi,\mathcal{C},\bar{\mathcal{P}},\xi,\bar\xi)$ is
completely  determined by the projection of $F$ on $\mathcal{M}$,
i.e., by the function $\bar{F}(\phi)=F(\phi,0,0,0,0,0)$, and a
function $\mathcal{O}(\phi)$ is the projection of some physical
observable iff
\begin{equation}\label{ob}
R_\alpha^i\partial_i\mathcal{O}=F_\alpha^bT_b
\end{equation}
for some $F_\alpha^b(\phi)$. Thus, to any on-shell gauge-invariant
function $\mathcal{O}$ on $\mathcal{M}$ one can associate a
BRST-invariant function on the extended phase space and vice
versa. Let $[F]$ denote the BRST-cohomology class of a physical
observable $F$, then  the map
\begin{equation}\label{exp-v}
[F]\mapsto \langle F\rangle_0 \equiv \bar F (\phi_0)\in
\mathbb{R}\,,
\end{equation}
with $\phi_0$ being a unique solution to Eqs.(\ref{t}),
establishes the isomorphism $\mathcal{H}_0\simeq \mathbb{R}$. By
definition, $\langle F \rangle_0$ is the classical (expectation)
value of the physical observable $F$.

One can also use the BRST language to give  another proof of the
classical equivalence of the topological theories associated with
the constraints $\Theta$ and $\tilde{\Theta}$ involving trivial
and nontrivial Lagrange anchors, respectively. Let $\Omega_1$
denote the BRST charge constructed by the former set of
constraints. Then the BRST charge (\ref{om}) is proved to be
canonically equivalent to $\Omega_1$. Namely, there exists a
function $G\in \mathcal{F}$ of ghost number zero such that
\begin{equation}\label{can-tr}
    \Omega=e^{\{G,\,\cdot\,\}}\Omega_1=\sum_{k=0}^\infty\frac1{k!}
    \{G,\{G,...\{G,\Omega_1\}...\}\,.
\end{equation}
The canonical equivalence of these two BRST charges implies the
isomorphism between corresponding cohomology groups, and hence the
physical equivalence of the classical theories defined by $\Omega$
and ${\Omega}_1$. In terms of the original constrained dynamics on
$T^\ast \mathcal{M}$, Rel.(\ref{can-tr}) implies the absence of
nontrivial deformations for the regular constraint systems
$\Theta$. In other words, any deformation  (\ref{deform}) is
obtained by a trivial superposition of a canonical transform of
$T^\ast \mathcal{M}$ and a linear combining of the initial
constraints, $\Theta_I\rightarrow G_I^J\Theta_J$, with some
nondegenerate matrix $G^J_I(\phi,\bar\phi)$.

\vspace{5mm}\noindent \textbf{Example 1.} Let $S(\phi)$ be a
nonsingular action functional, so that the corresponding
Van-Vleck's matrix $S_{ij}\equiv\partial_i\partial_jS$ is
invertible in a sufficiently small vicinity of a classical
solution $\phi_0$. The BRST charges corresponding to the zero and
the canonical anchors read
\begin{equation}\label{chs}
    \Omega_1=\bar{\eta}^i\partial_iS\,,\qquad \Omega =
    \bar\eta^i(\partial_iS-\bar\phi_i)\,.
\end{equation}

Let us show that the two BRST charges (\ref{chs}) are related to
each other by the canonical transform (\ref{can-tr}) with $G$
being a function of $\phi$ and $\bar\phi$. To this end, we first
split the phase-space variables onto the ``position coordinates''
$\varphi^I=(\phi^i,\eta_i)$ and their conjugate momenta
$\bar\varphi_J=(\bar\phi_j,\bar\eta^j)$,  and introduce an
auxiliary $\mathbb{N}$-grading counting the total number of
momenta entering monomials in $\bar\varphi$'s (the $m$-degree in
the terminology of Ref. \cite{KLS}). By definition,
$\mathcal{F}=\bigoplus_{k\in \mathbb{N}}\mathcal{F}_k$, where
$\mathcal{F}_k$ consists of homogeneous functions of $m$-degree
$k$:
\begin{equation}\label{}
    \mathcal{F}_k\ni F\,\Leftrightarrow\, NF=kF\,,\qquad
    N=\bar\varphi_I\frac {\partial}{\partial\bar\varphi_I}\,.
\end{equation}
Now we can prove the statement above by induction on the
$m$-degree. Observe that
\begin{equation}\label{G2}
\Omega^{(3)}\equiv e^{\{G_2,\,\cdot\,\}}\Omega_1 = \Omega +
O(\bar\varphi^3)\,,\qquad G_2=\frac 12
S^{ij}\bar\phi_i\bar\phi_j\,,
\end{equation}
$S^{ij}$ being the matrix inverse to $S_{ij}$. In other words, the
function $\Omega^{(3)}$ is canonically equivalent to $\Omega$
modulo $\bar\varphi^3$. Suppose $\Omega^{(k)}$ is canonically
equivalent to $\Omega$ modulo $k$, i.e.,
\begin{equation}\label{}
    \Omega^{(k)}=\Omega +\Omega_k + O(\bar\varphi^{k+1})\,, \qquad
    \Omega_k\in \mathcal{F}_k\,.
\end{equation}
It then  follows from the master equation
$\{\Omega^{(k)},\Omega^{(k)}\}=0$ that
\begin{equation}\label{}
\delta \Omega_k \equiv \{\Omega_1,\Omega_k\}=0\,,
\end{equation}
where we have introduced the nilpotent differential $\delta:
\mathcal{F}_k\rightarrow \mathcal{F}_{k}$ preserving $m$-degree.
Explicitly,
\begin{equation}\label{}
\delta=-\partial_iS\frac{\partial}{\partial\eta_i}+\bar\eta^iS_{ij}\frac{\partial}{\partial
\bar\phi_j}\,,\qquad \delta^2=0\,.
\end{equation}
Notice that the $\delta$-cohomology is nested in $\mathcal{F}_0$,
as we have the following contracting homotopy for $N$ with respect
to $\delta$:
\begin{equation}\label{}
\delta^\ast =\bar\phi_iS^{ij}\frac{\partial}{\partial \bar\eta^j}
\,,\qquad
    \delta\delta^\ast+\delta^\ast\delta = N\,.
\end{equation}
Then for any $k>0$ we have
\begin{equation}\label{Gk}
    \Omega_k=\delta G_{k}\,,\qquad G_{k}\equiv\frac1k
    \delta^\ast\Omega_k\,,
\end{equation}
and hence
\begin{equation}\label{}
\Omega^{(k+1)}\equiv e^{\{G_{k},\,\cdot\,\}}\Omega^{(k)} =\Omega
+O(\bar\varphi^{k+1})\,.
\end{equation}
Now we define the desired $G$ through the limit
\begin{equation}\label{lim}
    e^{\{G,\,\cdot\,\}}=\lim_{k\rightarrow \infty}
    e^{\{G_k,\,\cdot\,\}}\circ\cdots\circ e^{\{G_3,\,\cdot\,\}}\circ
    e^{\{G_2,\,\cdot\,\}}\,.
\end{equation}
By construction, $e^{\{G,\,\cdot\,\}}\Omega_1= \Omega$. We leave
it to the reader to check that the generator
\begin{equation}\label{}
G =\frac12
S^{ij}\bar\phi_i\bar\phi_j-\frac1{12}S^{kl}\partial_lS^{ij}\bar\phi_i\bar\phi_j\bar\phi_k+O(\bar\phi^4)\,,
\end{equation}
being defined by (\ref{G2}), (\ref{Gk}) and (\ref{lim}), does not
actually depend on $\eta$'s and $\bar\eta$'s. It may be shown that
the on-shell value of $G$, i.e., the function
$W_{tree}(\bar\phi)\equiv G(\phi_0,\bar\phi)$, coincides with the
generating function of connected Green's functions in tree
approximation.

\section{Probability amplitudes and a generalized Schwinger-Dyson equation}

In the previous section, a general gauge theory, whose equations
of motion (\ref{t}) are not necessarily Lagrangian, has been
equivalently reformulated as a constrained Hamiltonian system in
the phase space of fields and sources. Also the classical BFV-BRST
formalism has been constructed for this effective constrained
system that contains all the data concerning the original equations of
motion (\ref{t}), their gauge symmetries (\ref{RT}), and Noether
identities (\ref{ZT}). The physical observables of this effective
constrained system have been shown to be in one-to-one
correspondence with the on-shell gauge invariants of the original
non-Lagrangian theory. Now we are going to perform the operator
BFV-BRST quantization \cite{BFV}, \cite{BF1}, \cite{BF2},
\cite{HT} of this effective constrained system with a view to
studying physical states. By construction, the physical states of
this effective constrained system are the wave functions on the
(ghost-extended) space of all trajectories. Therefore
corresponding matrix elements of physical operators describe
quantum averaging over trajectories (histories) in the
configuration space of fields. In other words these matrix
elements are to be understood as the transition amplitudes of the
original non-Lagrangian theory. In standard Lagrangian field
theory these average values are usually described by Feynman's
path integral (\ref{Qaver}). Below we demonstrate how the above
definition of the transition amplitudes (in terms of the matrix
elements of the BFV-BRST quantized effective constrained dynamics)
reproduces the standard definition (\ref{Qaver}) in the Lagrangian
case. In the non-Lagrangian case, however, the corresponding
probability amplitude  $\Psi(\phi)$ cannot be brought to Feynman's
form (\ref{PA}) although it still defines a consistent quantum
dynamics.

To perform the operator quantization of the extended phase space
(\ref{pb}, \ref{PB}) in the Schr\"odinger  representation one
should first divide the phase-space variables into ``coordinates''
and ``momenta''. For our purposes it is convenient to choose them
as
\begin{equation}\label{phi}
    \varphi^I=(\phi^i,\eta_a,\xi_A,c^\alpha)\,,\qquad
    \bar\varphi_J=(\bar\phi_i,\bar\eta^a,\bar\xi^A, \bar c_\alpha)
\end{equation}
By construction, the expansion of the classical BRST charge
(\ref{om}) in powers of momenta contains no zeroth-order term:
\begin{equation}\label{exp}
    \Omega = \sum_{k=1}^\infty \Omega_k\,,\qquad \Omega_k=\Omega^{I_1\cdots I_k}(\varphi)\bar\varphi_{I_1}\cdots\bar\varphi_{I_k}\,.
\end{equation}
On substituting this expansion  into the classical master equation
(\ref{me}) we get a chain of equations
\begin{equation}\label{b}
    \{\Omega_1,\Omega_1\}=0\,,\quad
    \{\Omega_1,\Omega_2\}=0\,,\quad \{\Omega_2,\Omega_2\}= -2 \{\Omega_1,\Omega_3\} \,,\quad  \mathrm{etc}\,.
\end{equation}
As is seen,  the term linear in momenta, $\Omega_1$, is
Poisson-nilpotent by itself. In fact, $\Omega_1$ is nothing but
the ``bare'' BRST charge associated to the initial constraints
(\ref{ThetaInv}). According to (\ref{can-tr}), it is canonically
equivalent to the total BRST charge (\ref{exp}). Thus the leading
term of the BRST charge (\ref{exp}) carries all the information
about the classical gauge system (\ref{t}, \ref{ZT}, \ref{RT}),
with no reference to the Lagrange structure. The Lagrange anchor
$V$ enters $\Omega_2$, the term quadratic in momenta.  The second
equation in (\ref{b}), being expanded in ghost variables,
reproduces the defining relation for the Lagrange structure
(\ref{VT}). Notice that the sum $\Omega_1 + \Omega_2$ includes all
the ingredients of the Lagrange structure; the higher order terms
in (\ref{exp}) are added to get a Poisson-nilpotent function on
the extended phase space. It follows form Rels. (\ref{b}) that
$\Omega_2$ defines the so-called  \textit{weak antibracket} among
the momentum-independent functions:
\begin{equation}\label{abr}
(A,B)\equiv \{A,\{B,\Omega_2\}\}\,, \qquad \forall A(\varphi)\,,
B(\varphi)\,.
\end{equation}
In view of the third  relation in (\ref{b}) this antibracket
satisfies the Jacobi identity up to the homotopy associated with
the classical BRST-differential $ D A \equiv \{A, \Omega_1 \} $
(hence the name).  The second relation in (\ref{b})  implies that
$D$ differentiates the antibracket by the Leibniz rule. When the
antibracket is degenerate,  the operator $D$ may well be a
non-inner derivation of the anti-Poisson algebra, i.e., $D\neq
(W\,,\,\cdot\,)$ in general. For a more detailed discussion of the
$S_\infty$-structure behind this antibracket see \cite{KLS}.

Let us start quantizing this effective constrained system in the
Schr\"odinger representation, where the space of quantum states is
a complex Hilbert space of functions of $\varphi$'s w.r.t. the
hermitian inner product
\begin{equation}\label{inner}
\langle \Psi_1|\Psi_2\rangle =\int D\varphi
\Psi_1^\ast(\varphi)\Psi_2(\varphi)\,.
\end{equation}
The integration measure $D\varphi$ is given by the direct limit of
Lebesgue's measure upon the lattice approximation of field
configurations. The operators corresponding to the phase-space
variables act on the states by the rule
\begin{equation}\label{S-rep}
\hat{\varphi}^I\Psi=\varphi^I \Psi\,,\qquad
\hat{\bar\varphi}_I\Psi=-i\hbar\frac{\partial \Psi}{\partial
\varphi^I}\,,
\end{equation}
where all the derivatives act from the left. By definition,
\begin{equation}\label{}
[\hat{\bar\varphi}_I,\hat{\bar\varphi}_J]=0\,,\qquad
[\hat{\bar\varphi}_I,\hat{\varphi}^J]=-i\hbar\delta_I^J\,,\qquad
[\hat{\varphi}^I,\hat{\varphi}^J]=0\,.
\end{equation}
The hermiticity properties of the operators $\hat \varphi$ and
$\hat{\bar\varphi}$ follow directly from (\ref{hermprop}),
(\ref{phi}). The map
\begin{equation}\label{qp}
F(\varphi,\bar\varphi)=\sum_{k=0}^\infty F^{I_1\cdots
    I_k}(\varphi)\bar\varphi_{I_1}\cdots\bar\varphi_{I_k}\,
 \mapsto\,    \hat{F}=\sum_{k=0}^\infty
    F^{I_1\cdots I_k}(\hat{\varphi})\hat{\bar\varphi}_{I_1}\cdots
    \hat{\bar\varphi}_{I_k}
\end{equation}
assigns to any function $F\in \mathcal{F}$ a unique
pseudo-differential operator (the
$\hat{\varphi}\hat{\bar\varphi}$-ordering prescription is
applied). Identifying the elements of $\mathcal{F}$ with
$\varphi\bar\varphi$-symbols of operators, one can be free to add
to them any terms proportional to positive powers of $\hbar$
without any impact on the corresponding classical limit.

A crucial step in the operator BFV-BRST quantization \cite{BFV},
\cite{HT} is to assign a nilpotent operator $\hat{\Omega}$ to the
classical BRST charge (\ref{om}). The $\varphi\bar\varphi$-symbol
of the operator $\hat{\Omega}$  is to have  the form
\begin{equation}\label{symb}
    \Omega(\varphi,\bar\varphi, \hbar)=\sum_{k=0}^\infty
    \hbar^k\Omega^{(k)}(\varphi,\bar\varphi)\,,
\end{equation}
where the leading term $\Omega^{(0)}$ is given by  the classical
BRST charge (\ref{om}), and the higher orders in $\hbar$ are
determined by the requirements of hermiticity and nilpotency
\cite{BF2}:
\begin{equation}\label{hn}
    \hat{\Omega}^\dagger=\hat{\Omega}\,,\qquad \hat{\Omega}^2=0\,.
\end{equation}
It may well happen that no $\hat{\Omega} $ exists satisfying these
equations. In that case the classical theory admits no
self-consistent quantization (based on
$\varphi\bar\varphi$-symbols of operators). This is just the
phenomenon usually called the quantum anomaly. In what follows we
assume our theory to be anomaly free so that both equations
(\ref{hn}) hold true.

Furthermore, we assume that the $\varphi\bar\varphi$-symbol of the
quantum BRST charge (\ref{symb}) satisfies condition
\begin{equation}\label{flat}
    \Omega(\varphi, 0, \hbar)=0\,.
\end{equation}
In the theory we consider, this condition appears to be crucial
for existence of nontrivial BRST-invariant states, i.e., states
that are annihilated by the quantum BRST charge,
$\hat\Omega|\mathrm{\Psi}\rangle =0$. The trivial states are the
BRST-invariant states of the form $\hat{\Omega}|\Lambda\rangle$.
Taking the quotient of the space of BRST-invariant states by the
subspace of trivial ones gives the BRST-state cohomology.

The full BRST algebra includes, in addition to the nilpotent BRST
charge, the ghost-number operator:
\begin{equation}\label{}
    \hat{\mathcal{G}}=\frac
    i{2\hbar}\sum_I\gh(\varphi_I)[\hat{\varphi}^I\hat{\bar\varphi}_I+(-1)^{\epsilon(\varphi^I)}\hat{\bar\varphi}_I\hat{\varphi}^I]\,,
\qquad \hat{\mathcal{G}}^\dagger=-\hat{\mathcal{G}}\,,
\end{equation}
so that $ [\hat{\mathcal{G}}, \hat{F}]= \gh(\hat{F})\hat{F} $ for
any homogeneous $\hat F$. In particular,
\begin{equation}\label{g-om}
    [\hat{\mathcal{G}},\hat\Omega]=\hat \Omega\,.
\end{equation}

Under certain assumptions \cite{HT} the linear space of states
splits as a sum of eigenspaces of $\hat{\mathcal{G}}$ with
definite real ghost number. The physical states are usually
associated with the equivalence  classes of BRST-invariant states
at ghost number zero. A consistent consideration of physical
states in the Schr\"odinger representation is known to require
further enlargement of the extended phase space by the so-called
nonminimal variables \cite{HT}. These do not actually change the
physical content of the theory as one gauges them out by adding
appropriate terms to the original BRST charge. The nonminimal
sector just serves to bring the physical states to the
ghost-number zero subspace where one can endow them with a
well-defined inner product.

The standard nonminimal sector includes the Lagrange multipliers
$\lambda^I=(\lambda^a,\bar\lambda^\alpha)$ to the first class
constraints (\ref{ThetaInv}),  anti-ghosts
$\bar{\mathcal{C}}_I=(\bar\rho_a,\rho_\alpha)$, ghosts-of-ghosts
$(\sigma_A, \upsilon_A, \beta^A, \gamma^A)$ as well as their
conjugate momenta $\pi_I=(\bar\lambda_a, \lambda_\alpha)$,
$\mathcal{P}^I=(\rho^a,\bar\rho^\alpha)$, and $(\bar\sigma^A,
\bar\upsilon^A, \bar\beta_A, \bar\gamma_A)$. The ghost number
assignments of the new variables are given by
\begin{equation}
\begin{array}{lll}
    \gh(\lambda^I)=0\,, &\quad
    \gh(\bar{\mathcal{C}}^I)=-1\,,&\quad
    \gh(\sigma_A)=-1\,,\\[3mm]
 \gh(\upsilon_A)=0\,,&\quad
    \gh(\beta^A)=1\,,&\quad
    \gh(\gamma^A)=2\,.
    \end{array}
\end{equation}
The conjugate variables have the same parities but the opposite
ghost numbers; since all the constraints are bosonic, the Grassman
parity of any variable equals its ghost number modulo 2. We also
impose the reality conditions
\begin{equation}\label{}
   \pi^\ast=\pi\,,\qquad \bar{\mathcal{P}}^\ast=\bar{\mathcal{P}}\,,\qquad \bar\sigma^\ast=
\bar\sigma\,,\qquad
\upsilon^\ast=\upsilon\,,\qquad\bar\beta^\ast=\bar\beta\,,\qquad\gamma^\ast=\gamma\,,
\end{equation}
so that the canonically conjugate variables are either real or
pure imaginary depending on  their Grassman's parity.

\textit{From now on we will denote the classical BRST charge
(\ref{om}) defined in the minimal sector by $\Omega_\mathrm{min}$,
while $\Omega$ will stand for a total BRST charge}. The latter is
given by
\begin{equation}\label{tot}
    \Omega=\Omega_\mathrm{min}+\pi_I \mathcal{P}^I+
    \bar\beta_A\gamma^A+\upsilon_A\bar\sigma^A\,.
\end{equation}
 Clearly, ``putting hats'' on the r.h.s. of (\ref{tot}) yields a
hermitian and nilpotent operator $\hat{\Omega}$ whenever
$\hat{\Omega}_\mathrm{min}$ was so. To promote the Schr\"odinger
representation (\ref{S-rep}) to the fully extended phase space  we
place the variables $(\lambda^a, \lambda_\alpha, \rho^a,
\rho_\alpha, \sigma_A, \upsilon_A, \beta^A, \gamma^A)$ among the
coordinates $\varphi$ and consider the other nonminimal variables
as momenta belonging to the set $\bar\varphi$.

The physical states are defined as usual in the BRST theory:
\begin{equation}\label{of}
    \hat{\Omega}|\Phi\rangle=0\,,\qquad
\hat{\mathcal{G}}|\Phi\rangle =0\,.
\end{equation}
As the  BRST operator $\hat{\Omega}$ corresponds to the first
class constraints, being the deformed equations of motion and
gauge symmetry generators, Eqs.(\ref{of}) define a probability
amplitude on the ghost-extended space of trajectories. It is the
equation which is to be understood as Schwinger-Dyson equation for
the general (i.e., not necessarily Lagrangian) gauge theories.

The gauge system  at hands being a topological one (in the sense
discussed in the previous section), it might be naively expected
to have a 1-dimensional subspace of physical states spanned by a
unique (up to equivalence) ``vacuum'' state $|\Phi\rangle$. This
would be quite natural because the probability amplitude has to be
a unique function on the space of trajectories with prescribed
boundary conditions. Actually, this is not always the case in the
BRST theory: The physical dynamics may have several copies in the
BRST-state cohomology\footnote{A particular manifestation of this
phenomenon is known as \textit{doubling} \cite{HT}.}, and choosing
one of them is equivalent to imposing extra conditions on the
physical states over and above (\ref{of}) (see \cite{HM},
 \cite{MO}, \cite{BM}, \cite{MS}, \cite{FHP}, \cite{HT}). For
example, taking $|\Phi'\rangle$ to be annihilated by all the
momenta,
\begin{equation}\label{f'-con}
\begin{array}{l}
\hat{\bar\varphi}_I |\Phi'\rangle=0\,,
\end{array}
\end{equation}
we get a BRST-invariant state on account of (\ref{flat}). The
conditions (\ref{f'-con}) also ensure zero ghost number for
$|\Phi'\rangle$. Thus, $|\Phi'\rangle$ is a physical state.

A less trivial example of a physical state can be obtained by
imposing the following extra conditions:
\begin{equation}\label{f-con}
\begin{array}{l}
\hat{\eta}_a|\Phi\rangle=\hat{\bar{c}}_\alpha|\Phi\rangle=\hat{\xi}_A|\Phi\rangle=\hat{\bar\lambda}^\alpha
|\Phi\rangle
=\hat{\lambda}^a|\Phi\rangle=0\,,\\[3mm]
\hat{\rho}^a|\Phi\rangle=\hat{\bar\rho}^\alpha|\Phi\rangle=\hat{\gamma}^A|\Phi\rangle=\hat{\sigma}_A|\Phi\rangle=\hat{\beta}^A|\Phi\rangle=\hat{\upsilon}_A|\Phi\rangle=
0\,,
\end{array}
\end{equation}
Unlike $|\Phi'\rangle$, the form of the state $|\Phi\rangle$
essentially depends  on a particular structure of constraints. In
the Schr\"odinger representation (\ref{S-rep}) we have
\begin{equation}\label{ff'}
    \Phi(\varphi)=\delta(\eta)\delta(\xi)\delta(\lambda^a) \delta(\rho^a)\delta(\gamma)\delta(\sigma)\delta(\beta)\delta(\upsilon)\Psi(\phi)\,,\qquad
\Phi'(\varphi)=c\in \mathbb{C}\,,
\end{equation}
where the ``matter state'' $\Psi$ is annihilated by the quantum
constraint operators
\begin{equation}\label{thetapsi}
\hat{\Theta}_I\Psi=0 \quad\Leftrightarrow\quad\left\{%
\begin{array}{l}
(T_a(\phi) -i\hbar V^i_a(\phi)\partial_i+\cdots)\Psi(\phi)=0\,,
\\[3mm]
 (-i\hbar R_\alpha^i(\phi)\partial_i+\cdots)\Psi(\phi)=0\,. \\
\end{array}
\right.
\end{equation}
Of course,  the $\phi\bar\phi$-symbols of the quantum constraints
$\hat{\Theta}_I$ may differ from $\Theta_I(\phi,\bar\phi)$ by
$\hbar$-corrections defined by the quantum master equation
(\ref{me}). In the absence of gauge symmetry Eqs. (\ref{thetapsi})
reproduce the equation (\ref{GSDE}) for the (complex conjugate of)
probability amplitude of a (non-)Lagrangian field theory  we have
come to in Introduction on heuristical grounds. In that case the
states $|\Phi\rangle$ and $|\Phi'\rangle$ are just \textit{dual}
to each other \cite{HT}. Given the wave function
$\Phi^\ast(\varphi)$ - the probability amplitude on the extended
space of histories - we can calculate the quantum average of any
physical observable.

The physical observables are given by the BRST-cohomology classes
of hermitian operators at ghost number zero:
\begin{equation}\label{}
\begin{array}{c}
\hat{\mathcal{O}}^\dagger =\hat{\mathcal{O}}\,,\qquad
[\hat{\Omega},\hat{\mathcal{O}}]=0\,, \qquad
    [\hat{\mathcal{G}},\hat{\mathcal{O}}]=0\,,\\[3mm]
    \hat{\mathcal{O}}\sim \hat{\mathcal{O}}+[\hat{\Omega},
    \hat{B}]\,,\quad \gh(\hat{B})=-1\,.
    \end{array}
\end{equation}
If $\mathcal{O}(\varphi,\bar\varphi, \hbar)$ is the
$\varphi\bar\varphi$-symbol of an observable $\hat{\mathcal{O}}$,
then $\mathcal{O}(\varphi,\bar\varphi,0)$ is a classical
observable in the sense of definition (\ref{ob}). As with the BRST
charge, not any classical observable can be generally promoted to
the quantum level because of anomalies.

As is known, the inner product (\ref{inner}) becomes ill-defined
when restricted onto the physical subspace, and a regularization
is needed \cite{MS}, \cite{HT}. The usual receipt of regularizing
inner product of two physical states $|\Phi_1\rangle$ and
$|\Phi_2\rangle$ reads
\begin{equation}\label{in-pr}
    \langle \Phi_1|\Phi_2\rangle_K = \langle \Phi_1|e^{\frac i\hbar
    [\hat\Omega,\hat K]}|\Phi_2\rangle\,.
\end{equation}
Here $K$ is an appropriate gauge-fixing fermion of ghost number
$-1$. Formally, the expression (\ref{q-av}) does not depend on the
choice of $K$, since we fold the regulator $e^{[\hat\Omega,\hat
K]}$ between the states annihilated by $\hat\Omega$. (More
precisely, it depends only  on the homotopy class of $K$ in the
variety of all admissible gauge-fixing fermions.)

Now we postulate the number
\begin{equation}\label{q-av}
    \langle {\mathcal{O}}\rangle = \frac{\langle\Phi
    |\hat{\mathcal{O}}|\Phi'\rangle_K} {\langle\Phi|\Phi'\rangle_K}
\end{equation}
to be  the \textit{quantum average} of a physical observable
$\mathcal{O}(\phi)$ associated with the classical system (\ref{t})
and the Lagrange structure (\ref{VT}).

The reader may wonder why we sandwich the physical observable
between different copies of a single physical state, rather than
take the diagonal matrix elements
$\langle\Phi|\hat{\mathcal{O}}|\Phi\rangle_K$ or
$\langle\Phi'|\hat{\mathcal{O}}|\Phi'\rangle_K$ by analogy with
the definition of expectation values in conventional quantum
mechanics. The reasons are as follows:

(i) Upon normalization all the aforementioned expressions  give
the same value for the quantum average (see Proposition 2 of the
next section). In other words,  either copy of the physical state
is eligible.

(ii) It should be realized that given a physical observable
$\mathcal{O}$, expression (\ref{q-av}) defines actually an
infinite number of transition amplitudes associated with different
initial and final states of the original gauge theory (\ref{t})
even though it looks like a particular equal-time matrix element
of the effective Hamiltonian theory in $d+1$ dimensions. The
different initial and final states (w.r.t. the true physical time
containing among $d$ dimensions) enter implicitly through the
boundary conditions for the field $\phi$ in perfect analogy to the
usual Feynman's transition amplitudes.

(iii)  The special convenience of the ``asymmetric'' definition
(\ref{q-av}) is that it provides a direct link with the
conventional path-integral quantization in the case of Lagrangian
equations of motion (see Example 2 below). The relevance of
asymmetric inner products has been long known \cite{HM} for
tracing various relationships between the BRST theory and the
Dirac quantization.

Rel. (\ref{q-av}) has also the following ``cohomological''
interpretation similar to the definition of expectation values of
classical physical observables (\ref{exp-v}). To any physical
observable $\hat{\mathcal{O}}$ we can associate the physical state
$|O\rangle =\hat{\mathcal{O}}|\Phi'\rangle$. The BRST-state
cohomology being essentially one-dimensional, the state
$|O\rangle$ has to be proportional to  $|\Phi'\rangle$ up to a
BRST-trivial state\footnote{In principle, it may be proportional
to any copy of $|\Phi\rangle'$, say $|\Phi\rangle$, but a close
look at $|O\rangle$ rules out this possibility.}. So, we have
\begin{equation}\label{eigen}
    |O\rangle\equiv \hat{\mathcal{O}}|\Phi'\rangle = \langle
    \mathcal{O}\rangle |\Phi'\rangle + \hat\Omega|\Lambda\rangle\,,
\end{equation}
for some $\langle \mathcal{O}\rangle\in \mathbb{C}$ and
$|\Lambda\rangle$. Upon passing to the BRST-cohomology,
$|\Phi'\rangle$ becomes an eigenstate for any physical observable
$\hat{\mathcal{O}}$, and taking the inner product of (\ref{eigen})
with the BRST-invariant state $|\Phi\rangle$, we just identify the
eigenvalue $\langle \mathcal{O}\rangle$ with the quantum average
(\ref{q-av}) of $\hat{\mathcal{O}}$. Using the definition
(\ref{eigen}), we can rewrite (\ref{q-av}) in the form
\begin{equation}\label{q-av2}
    \langle {\mathcal{O}}\rangle = \frac{\langle \Phi|O\rangle_K} {\langle
    \Phi
    |1\rangle_K}= (\mathrm{const})\int D\varphi
    O(\varphi)\Phi^\ast_K(\varphi)\,,
\end{equation}
that  enables us to interpret the gauge-fixed probability
amplitude $ \Phi^\ast_K(\varphi)=\langle \Phi|\varphi\rangle_K$ as
a linear functional on the space of physical observables
represented by the states $O(\varphi)=\langle \varphi|O\rangle$.

 \vspace{5mm} \noindent \textbf{Example 2}. Consider a
Lagrangian gauge theory of rank 1, which is described by an action
$S(\phi)$ and a set of gauge algebra generators
$R_\alpha=R_\alpha^i\partial_i$ subject to the relations
\begin{equation}\label{SR}
    R_\alpha^i\partial_iS=0\,,\qquad
    [R_\alpha,R_\beta]=W_{\alpha\beta}^\gamma R_\gamma\,.
\end{equation}
The total and minimal BRST charges are given respectively by
\begin{equation}\label{tot1}
    \Omega=\Omega_\mathrm{min} + \bar\lambda_i {\rho}^i+  \lambda_\alpha
    \bar{\rho}^\alpha   + \bar\beta^\alpha{\gamma}_\alpha + {\upsilon}_\alpha \bar\sigma^\alpha
    \,,\qquad \Omega_{\mathrm{min}}=\Omega_1+\Omega_2\,,
\end{equation}
where
\begin{equation}\label{min}
\begin{array}{lll}
\Omega_1 &=& \bar\eta^i\partial_iS+c^\alpha
(R_\alpha^i\bar{\phi}_i - \eta_j\partial_iR_\alpha^j\bar\eta^i)+
c^\alpha\left(\frac12 c^\beta  W_{\beta\alpha}^\gamma\bar
c_\gamma-\xi_\gamma W_{\beta\alpha}^\gamma\bar\xi^\beta\right)\\[3mm]
&+&\frac12 c^\alpha c^\beta\partial_i W_{\beta\alpha}^\gamma
\xi_\gamma\bar\eta^i- \eta_iR_\alpha^i \bar\xi^\alpha\,,\\[3mm]
\Omega_2&=&\bar\eta^i\bar\phi_i+\bar c_\alpha\bar\xi^\alpha\,
\end{array}
\end{equation}
are terms linear and quadratic in momenta (cf. (\ref{exp})). In
the context of Lagrangian gauge theories the  truncated BRST
charge $\Omega_1$ was first introduced in \cite{GD}.  Quantizing
the classical BRST charge (\ref{tot1}) by the rule (\ref{qp}), one
gets an odd second-order differential operator
\begin{equation}\label{dif}
\hat\Omega=-i\hbar\left(\partial_jS(\phi)-i\hbar\frac{\partial}{\partial
\phi^j}\right)\frac{\partial}{\partial \eta_j}-i\hbar c^\alpha
R_\alpha^j(\phi)\frac{\partial}{\partial \phi^j}+\cdots
\end{equation}
acting on the Hilbert space of functions of
$$\varphi^I=(\phi^i, \eta_i, c^\alpha,\xi_\alpha,
\lambda^i,   \lambda_\alpha, \rho^i, \rho_\alpha, \sigma_\alpha,
\upsilon_\alpha, \beta^\alpha, \gamma^\alpha)\,.
$$
In general, this operator is neither hermitian nor nilpotent:
\begin{equation}\label{}
\hat{\Omega}^\dagger = \hat{\Omega} -2i\hbar
\hat{c}^\alpha\hat{A}_\alpha\,,\qquad
\hat{\Omega}^2=\hbar^2(\hat{A}_\alpha\hat{\bar\xi}^\alpha+\widehat{\partial
_iA}\hat{\bar\eta}^i)\,,
\end{equation}
here the function
\begin{equation}\label{anomaly}
A\equiv c^\alpha A_\alpha=c^\alpha(\partial_iR^i_\alpha
+W_{\alpha\beta}^\beta)
\end{equation}
represents  the so-called \textit{modular class} of gauge algebra
\cite{HC}, \cite{HT}. Taking the hermitian part of $\hat{\Omega}$,
we get a nilpotent BRST operator
\begin{equation}\label{O+A}
    \hat{\Omega}'=\frac12(\hat{\Omega}+\hat{\Omega}^\dagger)=\hat{\Omega}-{i\hbar}\hat{A}\,,\qquad
    \hat{\Omega}'\hat{\Omega}'=0\,.
\end{equation}
Occurrence of the second term in $\hat\Omega'$ conflicts, however,
with (\ref{flat}). As the result there may be no physical states
unless the modular class of $A$ is nontrivial. By construction,
$\{\Omega_1,A\}=0$, while the triviality means a stronger
condition  $A=\{\Omega_1, B\}$ for some $B(\varphi)$. In the
latter case one can pass to an equivalent quantization by twisting
the integration measure and the operator algebra,
\begin{equation} D'\varphi=D\varphi e^{2B}\,,\qquad \hat\varphi'{}^I=e^{-B}\hat\varphi^Ie^{
B}=\hat\varphi^I\,,\qquad \hat{\bar\varphi}'_I=e^{-
B}\hat{\bar\varphi}_Ie^{ B}=-i\hbar\partial_I-i\hbar\partial_IB\,,
\end{equation}
so that the $\varphi'\bar\varphi'$-symbol of the operator
(\ref{O+A}) will satisfy (\ref{flat}) modulo $\hbar^2$. The
nontrivial modular class indicates the presence of a genuine
1-loop anomaly. It should  be noted that in local field theory the
expressions like $A$ are singular and proportional to $\delta(0)$,
$\delta'(0)$, etc. It is beyond the scope of this paper to go into
details of regularizing quantum divergences, so we restrict our
consideration to the case $A=0$. Then the solution to the
generalized Schwinger-Dyson equation (\ref{of}) reads
\begin{equation}\label{sol}
\Phi(\varphi)=(\mathrm{const})\,\delta(\eta)\delta(\xi)\delta(\rho^i)\delta(\lambda^i)\delta(\gamma)\delta(\beta)\delta(\sigma)\delta(\upsilon)\,
e^{-\frac i\hbar S(\phi)}\,.
\end{equation}

Now let $\mathcal{O}(\phi)$ be a classical observable of the
original gauge theory (\ref{SR}), then
\begin{equation}
R_\alpha^i\partial_i\mathcal{O}=F_\alpha^i\partial_iS\,.
\end{equation}
The wave function corresponding to the state
$|O\rangle=\hat{\mathcal{O}}|\Phi'\rangle$ has the form
\begin{equation}\label{o-st}
O(\varphi)= \mathcal{O}(\phi)+c^\alpha F_\alpha^i(\phi)\eta_i
+\frac12 c^\beta c^\alpha K_{\alpha\beta}^\gamma(\phi)\xi_\gamma
+\cdots\,,
\end{equation}
where dots stand for higher powers of $\eta$'s and $\xi$'s.
Verifying  the BRST-invariance of $|O\rangle$, we find
\begin{equation}\label{o-an}
\hat{\Omega}{O}=\hbar^2\Lambda\,,\qquad \Lambda=c^\alpha
(\partial_iF^i_\alpha-K_{\alpha\beta}^\beta)+\cdots\,.
\end{equation}
By definition,  $\Lambda$ is a BRST-closed state with ghost number
1. Moreover, it is separately annihilated  by the
$\hat{\Omega}_1$- and $\hat\Omega_2$-parts of the total BRST
charge (\ref{tot1}). If $\Lambda$ is $\hat{\Omega}_1$-exact, i.e.,
$\hbar \Lambda=\hat{\Omega}_1{O}^{(1)}$, we can cancel the r.h.s.
of (\ref{o-an}) out by redefinition ${O}'={O}-\hbar {O}^{(1)}$, so
that the new state ${O}'$ will be BRST-invariant up to $\hbar^2$;
otherwise  no quantum observable corresponds to the classical
observable $\mathcal{O}$. Repeating this procedure time and again
we can move the anomaly at higher orders in $\hbar$ until we are faced
with a nontrivial $\hat{\Omega}_1$-cocycle at ghost number 1. In
that case the procedure stops and we get an unavoidable (or
genuine) quantum anomaly. Here we simply assume that $\Lambda=0$.

To regularize the inner product in the physical subspace spanned
by the state (\ref{sol}) and $\Phi'=1$  let us take the
gauge-fixing fermion
\begin{equation}\label{K}
    K=\frac i\hbar\rho_\alpha\chi^\alpha\,,
\end{equation}
where the functions $\chi^\alpha(\phi)$ are chosen in such a way
that the matrix $R_\alpha^i\partial_i\chi^\beta$ is nondegenerate.
Then we have
\begin{equation}\label{}
    [\hat{\Omega},\hat{K}]= c^\alpha
    R^i_\alpha\partial_i\chi^\beta\rho_\beta+\lambda_\beta\chi^\beta
    -i\hbar\partial_j\chi^\beta\rho_\beta\frac{\partial}{\partial\eta_j}\,.
\end{equation}
The gauge-fixed probability amplitude is given by
\begin{equation}\label{o-k}
\Phi^\ast_K(\varphi)=e^{-\frac i\hbar [\hat\Omega,\hat
K]}\Phi^\ast=
\delta(\eta_i-\partial_i\chi^\beta\rho_\beta)\delta(\xi)\delta(\lambda^a)\delta(\gamma)\delta(\beta)\delta(\sigma)\delta(\upsilon)e^{\frac
i\hbar (S-c^\alpha
R_\alpha^i\partial_i\chi^\beta\rho_\beta-\lambda_\beta\chi^\beta)}
\end{equation}
Inserting (\ref{o-st}) and (\ref{o-k}) into the general expression
(\ref{q-av2}) and integrating over arguments of
$\delta$-functions, we finally arrive at the Faddeev-Popov
path-integral
\begin{equation}\label{}
\langle \mathcal{O} \rangle =(\mathrm{const}){\int
{D}\phi{D}c{D}\rho{D}\lambda \,\bar{O} \exp{\frac
i\hbar(S-c^\alpha
R_\alpha^i\partial_i\chi^\beta\rho_\beta-\lambda_\alpha
\chi^\alpha)}}\,,
\end{equation}
where \begin{equation}
 \bar O (\phi,c,\rho)\equiv O(\phi, \eta, c, \xi
)|_{\xi=0, \,\eta_i=
\partial_i\chi^\beta\rho_\beta}=\mathcal{O}(\phi)+c^\alpha F(\phi)_\alpha^i\partial_i\chi^\beta\rho_\beta +
\cdots\,,
\end{equation}
and the normalization factor (const) is fixed by the condition
$\langle 1\rangle =1$.

\section{Path-integral representation for the probability amplitudes}

We have seen that the general solution to the generalized
Schwinger-Dyson equation (\ref{of}) and auxiliary conditions
(\ref{f-con}) is given by the product of $\delta$-functions of
ghosts and Lagrange multipliers and the matter state $\Psi(\phi)$
defined by the constraint equations (\ref{thetapsi}). The function
$\Psi^\ast(\phi)$ is then identified with the probability
amplitude on the original configuration space $\mathcal{M}$ of
fields with given boundary conditions. The gauge invariant
equations of motion give rise to the gauge invariant probability
amplitude $\Psi^\ast(\phi)$, much as a gauge invariant action
functional $S(\phi)$ leads to the gauge invariant Feynman's
amplitude $e^{\frac i\hbar S}$ in the Lagrangian field theory. In
both the cases an appropriate gauge-fixing procedure is required
to compute the quantum averages of physical observables. The
crucial difference, however, is that in the conventional BV
quantization the probability amplitude is known from the outset,
whereas in non-Lagrangian theory this amplitude is yet to be found
from the constraint equation (\ref{thetapsi}), which is hard to
solve in general. A perturbation solution can be obtained by
making use of a suitable path-integral representation that we are
proceeding to derive.

Let us start with the following two propositions.

 \vspace{5mm}
\noindent{\textbf{Proposition 1.}} Let $|\phi\rangle$ be a family
of physical states defined by the conditions
\begin{equation}\label{fi-con}
\begin{array}{c}
\hat\Gamma |\phi\rangle =0\,,\qquad
\hat{\phi}^i|\phi\rangle =\phi^i|\phi\rangle\,,\\[3mm]
    \Gamma= (\mathcal{C}^I, \bar\xi^A, \pi_I, \bar{\mathcal{C}_I},
\bar\sigma^A,\bar\upsilon^A,\bar\beta_A,\bar\gamma_A)\,,
\end{array}
\end{equation}
and $\Psi(\phi)$ be a solution of the constraint equation
(\ref{thetapsi}), then
\begin{equation}\label{psi-psi}
    \Psi(\phi_1)\Psi^\ast(\phi_0)=\langle\phi_1|\phi_0\rangle_K\,,
\end{equation}
for some admissible gauge-fixing fermion $K$.

The proof can be found in \cite{FHP}. Notice that the states
$|\phi\rangle$ are physically equivalent to each other and  have
well-defined inner product with the probability amplitude
(\ref{ff'}), (\ref{thetapsi}) on the extended configuration space,
\begin{equation}\label{}
    \langle \phi|\Phi\rangle =\Psi(\phi)\,.
\end{equation}
The last property characterizes the state $|\phi\rangle$ as
\textit{dual} to $|\Phi\rangle$.

\vspace{3mm} \noindent \textbf{Proposition 2}. The quantum average
(\ref{q-av}) of a physical observable $\mathcal{O}$ can be written
as
\begin{equation}\label{q-av1}
\langle\mathcal{O}\rangle =\frac{\langle\Phi'|\hat{\mathcal{O}}
|\Phi'\rangle_K}{{\langle\Phi'|\Phi'\rangle_K}}\,,
\end{equation}
where  $|\Phi'\rangle$ is the physical state (\ref{f'-con})
annihilated by all the momenta $\hat{\bar\varphi}$'s and $K$ is an
admissible gauge-fixing fermion.

Notice that unlike (\ref{q-av}) the alternative definition
(\ref{q-av1}) involves a single state $|\Phi'\rangle$ whose form
is known and does not depend on a particular structure of
constraints.

\vspace{3mm}\noindent

\noindent \textbf{Sketch of the proof.} The proof is based on
abelianization arguments. Locally, one can always find a canonical
transformation (\ref{can-tr}) bringing the total BRST charge
$\Omega$ into the form
\begin{equation}\label{abel}
    \Omega_1 = \Omega^I(\varphi)\bar\varphi_I=\bar\eta^aT_a + \bar\xi^AZ_A^a\eta_a + c^\alpha
    R_\alpha^i\bar\phi_i+
    \bar\lambda_a\rho^a+\lambda_\alpha\bar\rho^\alpha+ \bar\beta_A\gamma^A
    +\upsilon_A\bar\sigma^A\,,
\end{equation}
where the constraints $T$'s and $R$'s are in abelian involution,
\begin{equation}\label{ab-con}
    R_\alpha^i\partial_iT_a=0\,,\qquad [R_\alpha,R_\beta]=0\,,
\end{equation}
and $R_\alpha^i\partial_iZ_A^a=0$. This canonical transform is
then lifted to a unitary operator $\hat U$ such that
$\hat\Omega_1=\hat U\hat\Omega\hat U^\dagger$ and $\hat
U|\Phi'\rangle=|\Phi'\rangle$. (See \cite{BF1}, \cite{BF2} for the
proof of quantum abelization theorem).  Of course, the structure
functions $T$'s, $Z$'s and $R$'s (entering (\ref{abel})) may
differ from the original ones by quantum corrections that can be
obtained from (\ref{hn}).

For abelian constraints  we can take $\mathcal{O}$ to be strongly
annihilated by the gauge symmetry generators,
\begin{equation}\label{}
R_\alpha^i\partial_i\mathcal{O}=0\,.
\end{equation}
The function $\mathcal{O}(\phi)$ is thus BRST-invariant and does
not depend on the ghost variables and Lagrange multipliers. Now
consider the following gauge-fixing fermion:
\begin{equation}\label{}
    K=\frac i\hbar[\eta_a\lambda^a+\rho_\alpha
    \chi^\alpha+\xi_A\beta^A+\bar\gamma_AN^A_a\bar\eta^a+\lambda^aN_a^A\sigma_A]\,,\qquad
    \det(Z^a_BN^A_a)\neq 0\,.
\end{equation}
We have $[\hat \Omega_1, \hat K]=i\hbar \widehat{\{\Omega_1,K\}}$,
where
\begin{equation}\label{}
\begin{array}{c}
   i\hbar \{\Omega_1, K\} = (\lambda^aT_a+\rho^a\eta_a) + (\lambda_\alpha\chi^\alpha+ c^\alpha
    R_\alpha^i\partial_i\chi^\beta\rho_\beta)+(\rho^aN_a^A\sigma_A+\lambda^aN_a^A\upsilon_A)\\[3mm]
    +(\eta_aZ^a_A\beta^A+\xi_A\gamma^A)+(\bar\gamma_AN^A_aZ^a_B\bar\xi^B+\bar\beta_AN^A_a\bar\eta^a)\,.
    \end{array}
\end{equation}
The last sum contains noncommuting terms that somewhat complicates
calculations. Nonetheless, after some algebra one can find an
explicit expression for the state $e^{\frac i\hbar
[\hat\Omega,\hat K]}|\Phi'\rangle$. (We refer the reader to
\cite{MS}, where similar operatorial expressions were thoroughly
studied as well as more general gauges.) For example, choosing the
matrix $N_a^A$ in such a way that $N^A_aZ^a_B=\delta_B^A$, we find
\begin{equation}\label{eexp}
e^{\frac i\hbar [\hat\Omega,\hat K]}\Phi'(\varphi)= e^{\frac
i\hbar[(\lambda^aT_a+\rho^a\eta_a) + (\lambda_\alpha\chi^\alpha+
c^\alpha
    R_\alpha^i\partial_i\chi^\beta\rho_\beta)+(\rho^aN_a^A\sigma_A+\lambda^aN_a^A\upsilon_A)
    +c(\eta_aZ^a_A\beta^A+\xi_A\gamma^A)]}
\end{equation}
for some nonzero constant $c\in \mathbb{R}$. Substituting
(\ref{eexp}) into (\ref{q-av1}) and integrating w.r.t.
$\lambda^a$, $\lambda_\alpha$, $ \xi_A,\gamma^A$, $c^\alpha$,
$\rho_\alpha$, and $\beta^A$, we get
\begin{equation}\label{}
\begin{array}{c}
\langle\Phi'|\mathcal{\hat{O}}e^{\frac i\hbar[\hat\Omega,\hat
K]}|\Phi'\rangle =\\[3mm]
(\mathrm{const})\int D\phi D\eta D\upsilon D\sigma
\,\mathcal{O}(\phi)
\delta(T_a+N_a^A\upsilon_A)\delta(Z_A^a\eta_a)\delta(\eta_a+N_a^A\sigma_A)\delta(\chi)\det(R_\alpha^i\partial_i\chi^\beta)\,.
\end{array}
\end{equation}
Further integration of $\upsilon$, $\sigma$, and $\eta$ yields
\begin{equation}\label{oint}
\langle\mathcal{O}\rangle=(\mathrm{const}) \int
D\phi\,\mathcal{O}(\phi)\Psi(\phi)\mu(\phi)\,,
\end{equation}
where
\begin{equation}\label{}
    \Psi=\delta(T_{\bar a}(\phi))\mu_0(\phi) \,,\qquad \mu_0=\det(\delta^a_{\bar a},Z_A^a)\,,\qquad
    \mu=\delta(\chi)\det(R_\alpha^i\partial_i\chi^\beta)\,,
\end{equation}
$\{T_{\bar a}\}\subset\{T_a\}$ being a complete subset of
independent constraints. The overall constant in (\ref{oint}) is
determined by the normalization condition $\langle1\rangle=1$.

Clearly, the distribution $\Psi$ is nothing but the classical
probability amplitude annihilated by the operators of abelian
constraints (\ref{ab-con}). Substituting this $\Psi$ into
(\ref{ff'}) gives a solution to the generalized Schwinger-Dyson
equation (\ref{of}) with $\hat\Omega_1$ in place of $\hat \Omega$.
Explicitly,
\begin{equation}\label{f-ab}
 \Phi(\varphi)=\delta(\eta)\delta(\xi)\delta(\lambda^a)
 \delta(\rho^a)\delta(\gamma)\delta(\sigma)\delta(\beta)\delta(\upsilon)\mu_0(\phi)\delta(T_{\bar
 a})\,.
\end{equation}
The proof is completed by showing that the same expression
(\ref{oint}) for the quantum average is obtained if one starts
with the definition (\ref{q-av}), where $\Omega$ and $\Phi$ are
given respectively  by (\ref{abel}) and (\ref{f-ab}), and chooses
\begin{equation}\label{}
    K=\frac i\hbar \chi^\beta\rho_\beta
\end{equation}
as gauge-fixing fermion. The details are left to the reader.
$\square$

Regarding now the regulator $e^{\frac i\hbar[\hat\Omega,\hat K]}$,
which is involved in the definition  of inner products
(\ref{in-pr}), as the evolution operator associated to the
BRST-trivial Hamiltonian $[\hat\Omega,\hat K]$, one can
immediately get the path-integral representations for expressions
(\ref{psi-psi}) and (\ref{q-av1}). For the probability amplitude
we have
\begin{equation}\label{p-p}
    \Psi(\phi_1)\Psi^\ast(\phi_0)=\int \mathcal{D}\varphi
    \mathcal{D}\bar\varphi \exp{\frac i\hbar\int_0^1 dt (\bar\varphi_I\dot{\varphi}^I -
    \{\Omega,K\})}\,,
\end{equation}
where, in accordance with (\ref{fi-con}), integration extends over
all the paths obeying
\begin{equation}\label{}
\Gamma(0)=\Gamma(1)=0\,,\qquad
\phi^i(0)=\phi_0^i\,,\qquad\phi^i(1)=\phi_1^i\,.
\end{equation}
The other variables are unrestricted at the endpoints. Integrating
the function (\ref{p-p}) w.r.t. $\phi_0$ with an appropriate
weight we get a desired solution  of the constraint equations
(\ref{thetapsi}) as function of $\phi_1$.

Similarly, the path integral
\begin{equation}\label{}
\langle\mathcal{O}\rangle=\int \mathcal{D}\varphi
    \mathcal{D}\bar\varphi\, O(\varphi(1)) \exp{\frac i\hbar\int_0^1 dt (\bar\varphi_I\dot{\varphi}^I -
    \{\Omega,K\})}
\end{equation}
with the boundary conditions
\begin{equation}\label{}
    \bar\varphi_I(0)=\bar\varphi_I(1)=0
\end{equation}
gives the quantum average of a physical observable $\mathcal{O}$.
Here the value
\begin{equation}\label{}
    O(\varphi)=\mathcal{O}(\phi) + (ghost \;terms)
\end{equation}
is the $\bar\varphi$-independent part of a BRST-invariant
extension of $\mathcal{O}(\phi)$.  It is determined by the
equation
\begin{equation}\label{}
    \{\Omega_1, O\}=0\,,
\end{equation}
where $\Omega_1$ is the leading term of expansion  (\ref{exp}). In
Ref. \cite{KLS} this path-integral representation for quantum
averages was derived within a superfield approach to the
BV-quantization of topological sigma-models.

\section{Comparison with the field-antifield formalism}

To gain further insight into the quantization of (non-)Lagrangian
gauge theories described in the previous sections, it is
instructive to compare our method with the standard
field-anti-field formalism well-known for Lagrangian theories
\cite{BV}, \cite{HT}. We begin with a brief outline of the usual
BV-formalism in the form which is suitable for comparing with its
non-Lagrangian counterpart.

In the BV-formalism one starts with  a configuration space
$\mathcal{N}$ of fields $\phi^A$ endowed with an integration
measure $D\phi$. The set $\phi^A$ includes both physical and ghost
fields. The odd cotangent bundle $\Pi T^\ast \mathcal{N}$, with
$\phi^\ast_A$ being fiber coordinates, is known as the
field-anti-field supermanifold. Besides the Grassman parity, all
the fields $\phi^A$ and anti-fields $\phi^\ast_A$ carry definite
ghost numbers such that
$\mathrm{gh}(\phi^\ast_A)=-\mathrm{\gh}(\phi^A)-1$. For simplicity
sake, we assume that $\mathcal{N}$ is a super-domain with $D\phi$
being the canonical translation-invariant measure. The canonical
measure on $\mathcal{N}$ induces the canonical measure $D\phi
D\phi^\ast$ on the odd cotangent bundle $\Pi T^\ast \mathcal{N}$.
Using these data, one can define the so-called \textit{odd Laplace
operator}\footnote{The BV method can also be formulated with a
more general integration measure, see e.g. \cite{odd-lap}.}
\begin{equation}\label{}
    \Delta=(-1)^{\epsilon_A}\frac {\partial_l^2}{\partial \phi^A\partial
    \phi^\ast_A}\,,\qquad \Delta^2\equiv 0\,.
\end{equation}
By definition, the operator $\Delta$ is nilpotent and has ghost
number 1.

Let  $W(\phi,\phi^\ast)$ be an arbitrary even function of ghost
number zero.  Define the \textit{twisted} Laplace operator by the
rule
\begin{equation}\label{twist}
\hat{\Omega}=e^{-\frac{i}\hbar W}(-\hbar^2\Delta) \,e^{\frac
i\hbar W}=\hat{\Omega}_0+\hat{\Omega}_1+\hat{\Omega}_2\,,\qquad
\hat{\Omega}^2\equiv 0\,.
\end{equation}
Here
\begin{equation}\label{Wtwist}
    \hat{\Omega}_0=-i\hbar \Delta W+\frac 12
    (W,W)\,,\qquad \hat{\Omega}_1=-i\hbar(W,\,\cdot\,)\,,\qquad
    \hat{\Omega}_2=-\hbar^2\Delta\,,
    \end{equation}
and $ (\cdot , \cdot )$ stands for the \textit{antibracket} of two
functions,
\begin{equation}\label{a-br}
    (-1)^{\epsilon(A)}(A,B)\equiv \Delta(AB) - (\Delta A)B-(-1)^{\epsilon(A)}A\Delta
    B\,.
\end{equation}
Like $\Delta$, the operator $\hat{\Omega}$ is odd, nilpotent, and
has ghost number 1.  The difference between $-\hbar^2\Delta$ and
$\hat\Omega$ is in the first- and zeroth-order differential
operators $\hat\Omega_1$ and $\hat\Omega_0$ constructed from $W$,
see (\ref{Wtwist}).

The operator $\hat{\Omega}$ is said to be the \textit{quantum BRST
operator} associated to the \textit{quantum master action} $W$ if
\begin{equation}\label{master}
    \Omega_0 = 0\,.
\end{equation}
This condition is known as the \textit{quantum master equation}.
In the BV theory one is usually interested in \textit{proper}
solutions to the quantum master equation. These are specified by
two additional requirements: (i) $W$ is analytical in $\hbar$,
i.e., $W=W_0+\hbar W_1+\hbar^2 W_2+\cdots$, and (ii) the rank of
the Hesse matrix $d^2W_0$ equals $\dim \mathcal{N}$ at any point
of the stationary surface $dW_0=0$.  In particular, the latter
condition excludes the trivial solution $W=0$ whenever $\dim
\mathcal{N}\neq 0$.

Thus, to define a gauge dynamics on $\Pi T^\ast \mathcal{N}$ means
to specify a proper solution to Eq. (\ref{master}). The quantum
BRST operator in the BV theory can be understood as a twist
(\ref{Wtwist}) of the odd Laplacian with $W$ being the master
action.

Given a quantum BRST operator, the physical observables are
identified with zero ghost-number functions $O(\phi,\phi^\ast)$
obeying condition
\begin{equation}\label{oo}
    \hat{\Omega}{O}=0 \qquad \Leftrightarrow\qquad \Delta (O e^{\frac i\hbar
    W})=0\,.
\end{equation}
In view of (\ref{master}) the unit $1\in\mathbb{R}$ is
automatically a physical observable. Of a particular interest is
the following family of solutions to Eq. (\ref{oo}) :
\begin{equation}\label{pr-am}
    \Phi_K=\delta(\phi^\ast_A-\partial_AK)e^{-\frac i\hbar
    W}\,,\qquad \hat{\Omega}\Phi_K=0\,,
\end{equation}
$K(\phi)$ being an arbitrary odd function on $\mathcal{N}$ with
ghost number -1. The function $\Phi^\ast_K$ is nothing but the
gauge-fixed probability amplitude on the field-antifield
configuration space $\Pi T^\ast \mathcal{N}$. The quantum average
of a physical observables $O$ is defined now as
\begin{equation}\label{int}
\begin{array}{c}
\displaystyle\langle O\rangle =\int D\phi D\phi^\ast O \Phi^\ast_K
=\int D\phi D\phi^\ast \delta(\phi^\ast_A-\partial_AK)e^{\frac
i\hbar
    W} \\[5mm]
    \displaystyle = \int D \phi O(\phi,\partial K) \exp\frac i\hbar W(\phi,\partial
    K)\,.
\end{array}
\end{equation}
Because of (\ref{oo}) this integral does not actually depend on
$K$ (by Stokes theorem).

Notice that the BRST operator $\hat\Omega$ is hermitian w.r.t. the
canonical integration measure providing the antifields
$\phi^\ast_A$ are chosen to be pure imaginary.  Using this fact
and the definition (\ref{pr-am}) of the probability amplitude, one
can easily deduce the Word identity
\begin{equation}\label{}
    \langle O+\hat\Omega \Lambda \rangle =\langle
    O\rangle \quad \Leftrightarrow\quad \langle \hat\Omega \Lambda\rangle=0 \,,\qquad \forall \Lambda\,.
\end{equation}
In other words, the quantum average of a physical observable $O$
depends on the $\hat\Omega$-cohomology class $[O]$ rather than a
particular representative of this class.

The identification of Rels. (\ref{twist} -\ref{int}) with
analogous  constructions of Sect. 3 is now straightforward. The
quantum BRST operator (\ref{twist}) is just a particular example
of the general BRST charge defined by Rels. (\ref{exp},
\ref{symb}, \ref{hn}) where $\hat \Omega_k=0$, $\forall k > 2$,
and the matrix $\Omega^{IJ}$ determining $\hat{\Omega}_2$ is
constant and nondegenerate. The quantum master equation
(\ref{master}) corresponds to condition (\ref{flat}), the
\textit{flatness condition} in the terminology of Ref. \cite{KLS}.
The physical values (\ref{oo}) are naturally identified with the
BRST-invariant states (\ref{eigen}). In particular, the value
(\ref{pr-am}) - the complex conjugate to the gauge-fixed
probability amplitude - coincides in form with a particular
solution to the generalized Schwinger-Dyson equation (\ref{ff'}),
(\ref{o-k}). Finally, the following relation set up a
correspondence between minimal sectors of fields in a pure
Lagrangian case\footnote{As regards the non-minimal variables, one
can introduce them in many different, but equivalent, ways
\cite{HT}.}:
\begin{equation}\label{}
    \phi^A=(\phi^i, c^\alpha)\,,\qquad \phi^\ast_A = (\eta_i,
    \xi_\alpha)\,
\end{equation}
(see Example 2 of Sect. 3).  With all these identifications, one
can regard the BV path integral (\ref{int}) as the
quantum-mechanical matrix element (\ref{q-av2}), with the inner
product being the ordinary $L^2$-norm w.r.t. the canonical
integration measure on $\Pi T^\ast \mathcal{N}$.

As we have seen, the BV quantization scheme is a particular case
of the proposed method that works also in the theories which are
not Lagrangian. Now let us comment on the main properties of our
method which may have a more general form in non-Lagrangian case
in comparison with the Lagrangian BV scheme.

(i) In distinction from the canonical BRST operator (\ref{twist}),
the general BRST charge (\ref{exp}) can involve differential
operators of order $k > 2$ that gives rise to the \textit{higher
antibrackets} (see \cite{ABD}, \cite{V}, and references therein).

(ii) Even if all the higher antibrackets vanish (i.e., $\hat
\Omega_k=0$, $\forall k>2$), the quantum BRST operators
(\ref{twist}), (\ref{exp}) are not equivalent in general. This is
because the antibracket (\ref{abr}) associated with the quadratic
term in expansion (\ref{exp}) is allowed to be degenerate or even
irregular as distinct from the canonical antibracket (\ref{a-br}).
This fact has further consequences:

(iii) In the BV quantization, the first-order operator
$\hat\Omega_1=(W,\,\cdot\,)$ is, by definition, an inner
derivation of the canonical antibracket. For a degenerate
antibracket (\ref{abr}) this is not always the case: the
antibracket is still differentiated by the classical BRST
differential $\hat{\Omega}_1$, but this may well be a non-inner
derivation. In that case no quantum master action $W$ can be found
to twist the first order term $\hat \Omega_1$ out from the quantum
BRST operator $\hat\Omega =\hat\Omega_1+\hat\Omega_2$ by analogy
with (\ref{twist}). This also means that the quantum BRST operator
is not necessarily a twist (\ref{Wtwist}) of the odd Laplacian
$\hat\Omega_2$.

(iv) As a result, it turns out impossible to present the
probability amplitude $\Phi^\ast$ of a non-Lagrangian theory as
the exponential of some smooth function $e^{\frac{i}{\hbar}W}$
which is non-vanishing for every trajectory. The amplitude, being
defined by the generalized Schwinger-Dyson equation (\ref{of}),
can be a more general distribution on the space of all histories.
For example, the classical amplitude (\ref{ClA}) vanishes
everywhere except a classical solution.

As a final remark let us note that the proposed  method, being
applicable for covariant quantization of non-Lagrangian dynamics,
can be also useful for in-depth study of gauge anomalies in
conventional Lagrangian or Hamiltonian theories. Recall that in
the BV-quantization, the anomalies manifest themselves as
obstructions to solvability of the quantum master equation
(\ref{master}).  In particular, the 1-loop anomaly is given by the
modular class $[\Delta W_0]$ of the BRST-cohomology associated
with  the classical BRST-differential $\delta =(W_0,\,\cdot\,)$.
Notice that the $\delta$-cocycle $\Delta W_0$ has ghost number 1
and is linear in the gauge algebra generators $R$ as well as the
higher structure functions (see (\ref{anomaly})). Contrary to
this, in the operator BFV-BRST quantization the same anomaly
arises as violation of the nilpotency condition $\hat\Omega^2= 0$
by quantum corrections which are obviously bilinear in the
structure functions  and have ghost number 2. So it might be not
obvious in general how to relate these anomalies to each other.
Moreover, in the operator quantization the very existence of
anomalies depends crucially on the  quantization scheme applied,
not to mention their specific form. For instance, the 1-loop
anomalies may occur in the Wick quantization but never in the Weyl
one (see e.g. \cite{BF2}, \cite{HC}). What is a precise Lagrangian
analogue for different symbols of the BRST operator is yet to be
explored, but the proposed quantization scheme offers an
alternative way to handle this problem. The point is that the
gauge sector of the ``Lagrangian BRST charge'' (\ref{min}) is
quite similar to the Hamiltonian BRST charge constructed by the
constraints $R$'s, with the only difference that the former is
defined in the space of all histories while the latter corresponds
to a fixed time moment. Both of these BRST operators obey the same
master equation, $\hat\Omega^2=0$, leading to the same structure
of quantum anomalies. The different symbols of operators on the
genuine phase space of the  system are then imitated by different
symbols on the cotangent bundle of the space of all histories. In
summary, the proposed quantization scheme combines the explicit
covariance of the BV-quantization with the possibility to work
with different symbols of the  BRST operator, as is the case in
the BFV-BRST-quantization. Besides, the proposed method extends
these advantages beyond the reach of the BFV and BV schemes,
making possible to explore quantum effects, including anomalies,
in the theories admitting no Lagrangian formulation.

\end{document}